\documentclass[12pt, onecolumn, draftcls]{IEEEtran}
\usepackage{amsmath,amssymb}
\usepackage[dvips]{graphicx}
\usepackage{amsfonts}
\usepackage[mathscr]{eucal}
\usepackage{latexsym}
\usepackage{amsthm}
\usepackage{exscale}
\usepackage[mathscr]{eucal}
\usepackage{bm}
\usepackage[dvipsnames]{color}
\usepackage{cases}
\usepackage{epsfig}
\usepackage[center,small]{caption}
\usepackage{algorithm}
\usepackage{algorithmic}
\usepackage[verbose,nospace,sort]{cite}
\usepackage{tabularx}
\usepackage{multirow}
\usepackage{balance}
\scrollmode

\graphicspath{{./figs/}}

\IEEEoverridecommandlockouts

\newtheorem{theorem}{Theorem}
\newtheorem{remark}{Remark}
\newtheorem{lemma}{Lemma}

\newcommand{\xlog}{\mathrm{log}}

\newcommand{\bh}{\mathbf{h}}
\newcommand{\bx}{\mathbf{x}}

\newcommand{\bC}{\mathbf{C}}
\newcommand{\xC}{\widetilde{C}}
\newcommand{\xbC}{\widetilde{\mathbf{C}}}
\newcommand{\lbC}{\underline{\mathbf{C}}}

\newcommand{\xTr}{\mathrm{Tr}}

\begin{document}
\title{Optimized Transmission with Improper Gaussian Signaling in the $K$-User MISO Interference Channel}
\author{Yong~Zeng, Rui~Zhang, Erry~Gunawan,  and~Yong~Liang~Guan
\thanks{This paper has been presented in part at the International Conference on Acoustics, Speech, and Signal Processing (ICASSP), Vancouver, Canada, May 26-31 2013.}
\thanks{Y.~Zeng was with the School of Electrical and Electronic Engineering, Nanyang Technological University, Singapore. He is now with the Department of Electrical and Computer Engineering, National University of Singapore (email: elezeng@nus.edu.sg).}
 \thanks{R.~Zhang is with the Department of Electrical and Computer Engineering, National University of Singapore (e-mail: elezhang@nus.edu.sg).}
  \thanks{E.~Gunawan and Y.~L.~Guan are with the School of Electrical and Electronic Engineering, Nanyang Technological University, Singapore (e-mail: \{egunawan, eylguan\}@ntu.edu.sg).}
}

\maketitle

\begin{abstract}
This paper studies the achievable rate region of the $K$-user Gaussian multiple-input single-output interference channel (MISO-IC) with the interference treated as noise, when \emph{improper} or circularly \emph{asymmetric} complex Gaussian signaling is applied. The transmit optimization with improper Gaussian signaling involves not only the signal covariance matrix as in the  conventional \emph{proper} or circularly \emph{symmetric} Gaussian signaling, but also the signal \emph{pseudo-covariance} matrix, which is conventionally set to zero in proper Gaussian signaling. By exploiting the separable  rate expression with improper Gaussian signaling, we propose a separate transmit covariance and  pseudo-covariance optimization algorithm, which is guaranteed to improve the users' achievable rates over the  conventional proper Gaussian signaling. In particular, for the pseudo-covariance optimization, we establish the optimality of rank-1  pseudo-covariance matrices, given the optimal rank-1 transmit covariance matrices for achieving the Pareto boundary of the rate region. Based on this result, we are able to greatly reduce the number of variables in the  pseudo-covariance optimization problem  and thereby develop an efficient solution  by applying the celebrated semidefinite relaxation (SDR) technique. Finally, we extend the result to the Gaussian MISO broadcast channel (MISO-BC) with improper Gaussian signaling or so-called  widely linear transmit precoding.
\end{abstract}


\section{Introduction}
 The interference channel (IC) models the practical scenario in wireless communication when more than one transmitter-receiver pairs communicate independent messages simultaneously  over  the same frequency  and thus interfere with each other. Characterizing the information-theoretic capacity for the general $K$-user IC is a long-standing open problem  \cite{163}. 
  Moreover, since the capacity-approaching scheme in general requires multi-user encoding and decoding, which are difficult to implement in practical systems, a great deal of research on Gaussian ICs has focused on characterizing the achievable rate regions, under the assumption of employing single-user decoding (SUD) at receivers with the interference treated as noise. Among others, the achievable rate region has been characterized for  the Gaussian single-input single-output IC (SISO-IC) \cite{250}, single-input multiple-output IC (SIMO-IC) \cite{320}, and multiple-input single-output IC (MISO-IC) \cite{249,240,243,323}. In general, the achievable rate region of an IC is completely characterized by its Pareto boundary, which constitutes all the achievable rate-tuples for all users at each of which it is impossible to improve one user's rate without simultaneously decreasing the rate of at least one of the other users. One approach for such characterization is via solving a sequence of weighted-sum-rate maximization (WSRMax) problems. An alternative method based on the concept of ``rate profile'' was also proposed in  \cite{240} under the MISO-IC setup, which eventually results in solving a sequence of signal-to-interference-plus-noise ratio (SINR) feasibility problems that are easier to handle than WSRMax problems.  It is worth mentioning that for the MISO-IC, it has been shown by various methods in \cite{240,243,323} that all the rate-tuples on the Pareto boundary can be achieved with transmit beamforming, i.e., with rank-1 transmit covariance matrices.

Besides the achievable rate region characterization, significant research effort on Gaussian ICs has been devoted to solving the WSRMax problems \cite{320,253,325,233,313,356,335,321,322,345,326}. Unfortunately, such problems have been shown to be NP-hard in general \cite{325}. Many suboptimal algorithms have thus been proposed, e.g., the gradient descent algorithm \cite{233}, the interference-pricing based algorithm \cite{313}, the game-theory based algorithm \cite{356}, and the iterative weighted minimum mean-square-error (MMSE)  based algorithm \cite{335}. More recently, for Gaussian SISO-IC, SIMO-IC and MISO-IC, the globally optimal  solutions to WSRMax problems have been obtained  under the \emph{monotonic optimization} framework \cite{321,322,320,345,326}. However, the complexity of such globally optimal algorithms increases exponentially with the number of users, and their extension to the more general multiple-input multiple-output  IC (MIMO-IC) still remains unknown.  Moreover, it is worth mentioning that there have been a great deal of research interests over the last few years in characterizing  Gaussian ICs from the degrees-of-freedom (DoF) perspective using the technique of interference alignment (IA) (see \cite{348} and the references therein).

  The existing  works on ICs have mostly assumed \emph{proper} or circularly \emph{symmetric} complex Gaussian (CSCG) signaling for transmitted signals. It is worth noting that the more general \emph{improper} or circularly \emph{asymmetric} complex signal processing has  been exploited before in various other areas such as  widely linear estimation \cite{269}, while only recently it was revealed that improper complex Gaussian signaling, jointly applied with the techniques of  symbol extension and IA, is able to improve the achievable sum-rate DoF for the  three-user  Gaussian SISO-IC at the asymptotically high signal-to-noise ratio (SNR) \cite{189}. Later, it was shown that even for the two-user SISO-IC where IA is not applicable, the achievable rate region can still be enlarged with improper Gaussian signaling over the conventional proper Gaussian signaling at finite SNR \cite{255,256}. Furthermore, it has been shown  in our prior work \cite{350} that with improper Gaussian signaling, the user's achievable rate in the general MIMO-IC can be expressed as the summation of the rate achievable  by the conventional proper Gaussian signaling, which depends on the users' transmit covariance matrices only, and an additional term, which is a function of both the users' transmit covariance and \emph{pseudo-covariance} matrices. Such a separable rate structure has been exploited in \cite{350} to optimize the covariance and pseudo-covariance separately so that the obtained improper Gaussian signaling strictly outperforms the conventional proper Gaussian signaling in terms of the achievable rate region. However, the algorithms proposed in \cite{350} are for the two-user SISO-IC and cannot be applied when there are more than two users and/or multiple  antennas at the transmitter. This thus motivates our current work that extends the result in \cite{350} to optimize the transmission with improper Gaussian signaling in  the more general $K$-user MISO-IC.

Similar to \cite{240}, in this paper we apply the rate-profile technique to characterize the achievable rate region of the MISO-IC with the interference treated as Gaussian noise.  However, unlike the case with proper Gaussian signaling considered in \cite{240}, the optimization problem with improper Gaussian signaling  is non-convex and thus difficult to be solved optimally.  By adopting the similar separate covariance and pseudo-covariance optimization approach as in \cite{350}, we develop an efficient solution for this problem in the MISO-IC case. Specifically, for the pseudo-covariance  optimization, we first establish the optimality for rank-1 pseudo-covariance matrices given the optimal rank-1 transmit covariance matrices for achieving the Pareto optimal rates. Moreover, we show that each rank-1 pseudo-covariance matrix is parameterized by one unknown complex scalar. Based on this result, we formulate the original matrix optimization problem to an equivalent vector optimization problem in  considerably  lower dimensions. We then apply the celebrated semidefinite relaxation (SDR) technique \cite{349} to find an efficient approximate solution for the reformulated problem. It is worth noting that the approach of using SDR for solving non-convex quadratically constrained quadratic programs (QCQPs) has been successfully applied to find high-quality approximate solutions for various problems in communications and signal processing (see \cite{349} and the references therein). For our pseudo-covariance optimization problem under the $K$-user MISO-IC setup, we show that the proposed SDR-based solution is in fact optimal when $K=2$.  Finally, we show that the improper Gaussian signaling scheme developed for the MISO-IC  can also be  applied to the $K$-user MISO broadcast channel (MISO-BC), under the assumption of employing widely linear precoding  at the transmitter.\footnote{It is worth noting that if the optimal non-linear ``dirty paper coding (DPC)'' based precoding is applied for the MISO-BC, the achievable rate region can be equivalently characterized in the dual uplink SIMO multiple-access channel (MAC) via the celebrated BC-MAC duality result (see \cite{212} and the references therein), from  which it can be shown that proper Gaussian signaling is optimal for MISO-IC with DPC based nonlinear precoding.}

The rest of this paper is organized as follows. Section~\ref{S:systemModel} presents the MISO-IC model and the  problem formulation. Section~\ref{S:separate} develops the separate covariance and pseudo-covariance optimization algorithm for our formulated problem. In Section~\ref{S:MISO-BC}, the proposed improper Gaussian signaling scheme is extended to the  MISO-BC  with widely linear precoding.  Section~\ref{S:simulation} presents numerical results. Finally, we conclude the paper in Section~\ref{S:conclusion}.

\emph{Notations:} In this paper, scalars are denoted by italic letters. Boldface lower- and upper-case letters denote vectors and matrices, respectively. For a square matrix $\mathbf{S}$, $\mathrm{Tr}(\mathbf{S})$ denotes the trace. $\mathbf{S}\succeq \mathbf{0}$ and $\mathbf{S}\succ \mathbf{0}$ mean that $\mathbf{S}$ is positive semidefinite and positive definite, respectively. $\mathbb{C}^{M\times N}$ and $\mathbb{R}^{M\times N}$  denote the space of $M\times N$ complex and real matrices, respectively. For an arbitrary matrix $\mathbf{A}$, $\mathbf{A}^*$, $\mathbf{A}^{T}$, $\mathbf{A}^{H}$ and $\mathrm{rank}(\mathbf{A})$ represent the complex-conjugate, transpose, Hermitian transpose and rank of $\mathbf{A}$, respectively. The symbol $i$ denotes the imaginary unit, i.e., $i^2=-1$. $[\mathbf v]_j$ denotes the $j$th element of the vector $\mathbf v$, while $\|\mathbf v\|$ denotes its Euclidean norm. For a complex number $x$, $|x|$ denotes its magnitude. $\mathcal{CN}(\mathbf{x},\mathbf{\Sigma})$ represents the CSCG distribution of a random vector (RV) with mean $\mathbf{x}$ and covariance matrix $\mathbf{\Sigma}$. $\Re\{\cdot\}$ and $\Im \{\cdot\}$ represent the real and imaginary parts of a complex number, respectively. Finally, $\log(\cdot)$ denotes the logarithm function with base $2$.

\section{System Model and Problem Formulation}\label{S:systemModel}
We consider a $K$-user MISO-IC, where each transmitter is intended to send independent information to its corresponding receiver, while possibly interfering with all other $K-1$ receivers.\footnote{The techniques developed  in this paper can also be applied  to other interference-limited wireless systems, such as the multi-cell network with various levels of base station cooperation \cite{381}. In this paper, we will mainly focus on the MISO-IC  and briefly discuss its extension to the MISO-BC in Section~\ref{S:MISO-BC}.} Assume that each transmitter is equipped with $M>1$ antennas and each receiver with one antenna. The received baseband signal for user $k$ is given by
\begin{align}\label{E:yk}
y_k(n)=\mathbf h_{kk} \mathbf x_k(n) +\sum_{j\neq k} \mathbf h_{kj} \bx_j(n) + v_k(n), \ \forall k,
\end{align}
where $n$ is the symbol index; $\mathbf h_{kk} \in \mathbb{C}^{1\times M}$ denotes the direct channel  from transmitter $k$ to receiver $k$, while $\mathbf h_{kj}\in \mathbb{C}^{1\times M}, j\neq k$, denotes the interference channel from transmitter $j$ to receiver $k$; we assume quasi-static fading and thus all channels are constant over $n$'s for the case of our interest; $v_k(n)$ represents the independent and identically distributed (i.i.d.) zero-mean CSCG noise with variance $\sigma^2$, i.e., $v_k(n)\sim \mathcal{CN}(0, \sigma^2)$; and $\bx_k(n)\in \mathbb{C}^{M\times 1}$ is the transmitted signal vector from transmitter $k$, which is independent of $\mathbf x_j(n)$ for $j\neq k$. In this paper, for the purpose of exposition, we assume that the technique of symbol extensions over time \cite{189} is not used. Hence, $\mathbf x_k(n)$ is independent over $n$. For brevity, $n$ is omitted in the rest of this paper. Different from the conventional schemes where proper Gaussian signaling is assumed, i.e., $\bx_k\sim \mathcal{CN}(\mathbf 0, \bC_{\bx_k})$, with $\bC_{\bx_k}$ denoting the transmit covariance matrix, in this paper, we consider the more general improper Gaussian transmitted signals. For the background knowledge of improper (Gaussian) RVs, the readers may refer to \cite{350,357} and the references therein.

For the zero-mean transmitted Gaussian RV $\mathbf x_k$, we denote its covariance and pseudo-covariance matrices as $\bC_{\bx_k}$ and $\xbC_{\bx_k}$, respectively, i.e.,
\begin{align}
\bC_{\bx_k}= \mathbb{E} (\bx_k \bx_k^H),\ \xbC_{\bx_k}= \mathbb{E} (\bx_k \bx_k^T),
\end{align}
where $\bC_{\bx_k}$ is Hermitian and positive semidefinite, and $\xbC_{\bx_k}$ is symmetric in general. For the conventional proper Gaussian signaling, the pseudo-covariance matrices $\xbC_{\bx_k}$'s  for all transmitters are set to zero matrices, and thus are not included in  the transmit optimization. However, for the more general improper Gaussian signaling considered in this paper, the additional degrees of freedom given by the pseudo-covariance matrices provide a further  opportunity for improving rate over proper Gaussian signaling \cite{348}. $\bC_{\bx_k}$ and $\xbC_{\bx_k}$ are a valid pair of covariance and pseudo-covariance matrices, i.e., there exists a RV $\mathbf x_k$ with covariance and pseudo-covariance matrices given by $\bC_{\bx_k}$ and $\xbC_{\bx_k}$, respectively,  if and only if the corresponding augmented covariance matrix $\lbC_{\mathbf \bx_k}$ defined below is positive semidefinite \cite{246},
\begin{align}\label{E:validPair}
\lbC_{\mathbf \bx_k}\triangleq \bigg[\begin{matrix}\bC_{\mathbf{\bx_k}} & \xbC_{\bx_k} \\ \xbC_{\bx_k}^* & \bC_{\bx_k}^* \end{matrix} \bigg]\succeq \mathbf 0.
\end{align}
 For the MISO-IC, the  covariance and pseudo-covariance of the received signal $y_k$ can be written as
\begin{align}
C_{y_k}&= \mathbb{E} (y_k y_k^*)=\sum_{j=1}^K \mathbf h_{kj}\bC_{\bx_j} \mathbf h_{kj}^H+\sigma^2, \label{E:MISO_Cyk}\\
\xC_{y_k}&= \mathbb{E} (y_k y_k)=\sum_{j=1}^K \mathbf h_{kj} \xbC_{\bx_j} \mathbf h_{kj}^T. \label{E:MISO_PCyk}
\end{align}
It can be seen that $C_{y_k}$ is a positive real number whose value equals to the total received power at receiver $k$.
Denote the interference-plus-noise term at receiver $k$ as $s_k$, i.e.,
$
s_k=\sum_{j\neq k} \mathbf h_{kj} \bx_j + v_k.
$
Then the covariance and pseudo-covariance of $s_k$ are given by
\begin{align}
C_{s_k}&= \mathbb{E} (s_k s_k^*)=\sum_{j\neq k} \mathbf h_{kj} \bC_{\bx_j}\mathbf h_{kj}^H +\sigma^2,\\
\xC_{s_k}&= \mathbb{E} (s_k s_k)=\sum_{j\neq k} \mathbf h_{kj} \xbC_{\bx_j} \mathbf h_{kj}^T.\label{E:MISO_PCsk}
\end{align}
With improper Gaussian signaling at all  transmitters, the resulting interference at each receiver is improper Gaussian as well. Under the assumption that the improper Gaussian interference is treated as additional noise over existing proper Gaussian background noise at all receivers, an achievable rate expression has been derived in \cite{350} for the $K$-user MIMO-IC. By applying the result in  \cite{350} to the MISO-IC setup,  the achievable rate of user $k$ can be expressed as
\begin{align}
R_k&=\underbrace{\xlog\left(1+  \frac{\bh_{kk}\bC_{\bx_k}\bh_{kk}^H}{\sigma^2+\sum_{j\neq k} \bh_{kj}\bC_{\bx_j}\bh_{kj}^H} \right)}
 _{\triangleq R_{k}^{\text{proper}}(\{\bC_{\bx_j}\})}\notag \\
&\qquad \qquad \qquad \qquad  \qquad+\frac{1}{2}\xlog\frac{1-C_{y_k}^{-2}|\xC_{y_k}|^2}{1-C_{s_k}^{-2}|\xC_{s_k}|^2}.\label{E:Rk}
\end{align}
It is observed from \eqref{E:Rk} that with improper Gaussian signaling, each user's achievable rate is a summation of the rate achievable by the conventional proper Gaussian signaling, denoted by $R_{k}^{\text{proper}}(\{\bC_{\bx_j}\})$, and an additional term, which is a function of both the users' covariance and pseudo-covariance matrices. Therefore, for a given set of transmit covariance matrices obtained by any proper Gaussian signaling scheme, the achievable rates in MISO-IC can be improved with improper Gaussian signaling by choosing the pseudo-covariance matrices that make the second term in \eqref{E:Rk} strictly positive.

\begin{remark}
From user $k$'s perspective, with improper Gaussian signaling employed by other transmitters  and under the assumption that the resulting interference is treated as additional noise over existing proper or CSCG background noise at the receiver, user $k$ essentially communicates over a point-to-point MISO channel corrupted by \emph{improper} or circularly \emph{asymmetric} complex Gaussian noise. Therefore, the well-known result that proper Gaussian signaling is optimal \cite{36}, which is applicable for point-to-point channels with proper or CSCG noise only,  does not apply here. This thus motivates our work on investigating the more general improper Gaussian signaling for MISO-IC.
\end{remark}

The achievable rate region $\mathcal R$ for the $K$-user MISO-IC is defined as the set of rate-tuples that can be simultaneously achieved by all users under a given set of transmit power constraints at each transmitter, denoted by $P_k,k=1,\cdots,K$. With $R_k$ given in  \eqref{E:Rk}, we thus have
  \begin{align}
\mathcal{R}\triangleq \bigcup_{\begin{subarray}{l} \xTr(\bC_{\bx_k})\leq P_k, \\ \lbC_{\bx_k}\succeq \mathbf{0},\  \forall k\end{subarray}} \bigg\{(r_1, \cdots, r_K): 0\leq r_k \leq R_k,\forall k\bigg\},\label{E:region}
\end{align}
where the constraint $\lbC_{\bx_k}\succeq \mathbf{0}$ follows from \eqref{E:validPair}.

To characterize the Pareto boundary of the achievable rate region $\mathcal{R}$, we adopt the rate-profile method as in \cite{240}.  Specifically, any Pareto-optimal rate-tuple on the boundary can be obtained by solving the following optimization problem with a given rate-profile vector denoted by {\boldmath$\alpha$}$=(\alpha_1\cdots \alpha_K)$.
\begin{align}
\text{(P1):}    &  \underset{\{\bC_{\bx_k}\},\{\xbC_{\bx_k}\}, R}{{\mathrm{max.}}}   R \notag \\
 &  \qquad \ \text{s.t.} \quad     R_{k}\geq \alpha_k R, \ \forall k, \notag \\
& \qquad \qquad \ \ \xTr(\bC_{\bx_k})\leq P_k, \ \forall k,\notag \\
& \qquad \qquad \ \ \bigg[\begin{matrix}\bC_{\bx_k} & \xbC_{\bx_k} \\ \xbC_{\bx_k}^* & \bC_{\bx_k}^* \end{matrix} \bigg]\succeq \mathbf{0}, \ \forall k,\notag
\end{align}
where $\alpha_k$ denotes the target ratio between user $k$'s achievable rate and the users' sum-rate, $R$. Without loss of generality, we assume  $\alpha_k> 0, \forall k$, and $\sum_{k=1}^K \alpha_k=1$. Denote the optimal value of (P1) as $R^{\star}$. Then the rate-tuple $R^{\star}\cdot$~{\boldmath$\alpha$} must be on the Pareto boundary of the rate region $\mathcal{R}$. Thereby, by solving (P1) with different rate-profile vectors {\boldmath$\alpha$}, the complete Pareto boundary of $\mathcal{R}$ can be found. 

\section{Separate Covariance and Pseudo-Covariance Optimization}\label{S:separate}
(P1) is a non-convex optimization problem, and thus it is difficult to achieve its  global optimum  efficiently. In this section, we propose a separate covariance and pseudo-covariance optimization algorithm by utilizing  the separable rate expression given in \eqref{E:Rk} to obtain an efficient suboptimal solution for (P1). Specifically, the covariance matrices of the transmitted signals are first optimized by setting  the pseudo-covariance matrices to zero, i.e., proper Gaussian signaling is applied. Then, the pseudo-covariance matrices are optimized with the covariance matrices fixed as the previously obtained solution. With such a separate optimization approach, the obtained improper Gaussian signaling design is guaranteed to improve the achievable rates over the conventional proper Gaussian signaling.

\subsection{Covariance Optimization}\label{S:covarianceOpt}
When restricted to proper Gaussian signaling by setting $\xbC_{\bx_k}=\mathbf{0}, \forall k$, the second term in \eqref{E:Rk} vanishes to zero and (P1) reduces to
\begin{align}
 \text{(P1.1):} \ &  \underset{r,\{\bC_{\bx_k}\}}{{\mathrm{max.}}}  \  r \notag \\
    & \quad  \text{s.t.}  \     \xlog\left(1+\frac{\bh_{kk}\bC_{\bx_k}\bh_{kk}^H}{\sigma^2+\sum_{j\neq k} \bh_{kj}\bC_{\bx_j}\bh_{kj}^H} \right)\geq  \alpha_k r, \ \forall k, \notag \\ 
& \qquad \ \  \xTr(\bC_{\bx_k})\leq P_k,  \ \forall k, \notag \\
&\qquad \ \ \bC_{\bx_k}\succeq \mathbf{0}, \ \forall k.\notag
\end{align}
 Denote the optimal value of (P1.1) as $r^{\star}$. Then the rate-tuple  $r^{\star}\cdot \boldsymbol \alpha$ is on the Pareto boundary of the achievable rate region with proper Gaussian signaling.  It has been shown in  \cite{240,243,323} that all the Pareto-optimal rate-tuples with proper Gaussian signaling can be achieved by  rank-1 transmit covariance matrices or beamforming. Therefore, without loss of optimality for  (P1.1), we can let
\begin{align}
\bC_{\bx_k}=\mathbf{t}_k\mathbf t_k^H, \ \forall k,
 \end{align}
 where $\mathbf t_k$ is the transmit beamforming vector for user $k$.
 Then for any fixed target rate $r$, the feasibility problem related to (P1.1) can be formulated as
\begin{align}
\text{(P1.2):}  \ &  \text{Find}   \   \{\mathbf t_k\} \notag \\
\text{s.t.}  \ & \sigma^2+\sum_{j\neq k}|\bh_{kj}\mathbf t_j|^2 \leq \frac{1}{2^{\alpha_k r}-1} (\bh_{kk}\mathbf{t}_k)^2, \ \forall k,\notag\\
& \Im\{\mathbf h_{kk}\mathbf t_k\}=0, \  \|\mathbf t_k\|^2\leq P_k, \ \forall k, \notag
\end{align}
where without loss of generality, we have assumed that for each user $k$, $\mathbf h_{kk}\mathbf t_k$ is a nonnegative real number \cite{214}. (P1.2) is a second-order cone programming (SOCP) problem, which can be efficiently solved  \cite{202}. If (P1.2) is feasible, then the optimal value of (P1.1) satisfies $r^{\star}\geq r$; otherwise, $r^{\star}<r$. Therefore, (P1.1) can be optimally solved by solving (P1.2) with different values of  $r$, and  applying a  bisection search over $r$ \cite{202}.

\subsection{Pseudo-Covariance Optimization}\label{S:pseudoOpt}
Denote the optimal solution to (P1.1) as $\{r^{\star}, \bC_{\bx_k}^{\star}=\mathbf t_k\mathbf t_k^{H}\}$. By fixing the transmit covariance matrices as $\{\bC_{\bx_k}^{\star}=\mathbf t_k\mathbf t_k^{H}\}$, (P1) is further optimized over the pseudo-covariance matrices $\{\xbC_{\bx_k}\}$ in this subsection. By replacing the first term in the rate expression \eqref{E:Rk} by $\alpha_k r^{\star}$, the problem for pseudo-covariance matrix optimization is  formulated as
\begin{align}
 \text{(P1.3):} \   \underset{R, \{\xbC_{\bx_k}\}}{\mathrm{max.}} &  \quad  R \notag \\
 \text{s.t.} \ &  \alpha_k r^{\star}+\frac{1}{2}\xlog\frac{1-C_{y_k}^{-2}|\xC_{y_k}|^2}{1-C_{s_k}^{-2}|\xC_{s_k}|^2}\geq \alpha_k R , \ \forall k, \notag \\
&\bigg[\begin{matrix}\mathbf t_k\mathbf t_k^{H} & \xbC_{\bx_k} \\ \xbC_{\bx_k}^* & (\mathbf t_k \mathbf t_k^{H})^* \end{matrix} \bigg]\succeq \mathbf{0}, \ \forall k,
\end{align}
where $C_{y_k}$ and $C_{s_k}$ are fixed covariances given the previously optimized  transmit covariance matrices $\{\bC_{\bx_k}^{\star}=\mathbf t_k\mathbf t_k^{H}\}$; $\xC_{y_k}$ and $\xC_{s_k}$ are the pseudo-covariances, which are related to  transmit pseudo-covariance matrices $\{\xbC_{\bx_k}\}$ via \eqref{E:MISO_PCyk} and \eqref{E:MISO_PCsk}, respectively.    By treating $R$ as a slack variable and discarding irrelevant terms in (P1.3), the problem can be re-written as a minimum-weighted-rate maximization (MinWR-Max) problem as follows.
 \begin{align}
 \text{(P1.4):} \    \underset{\{\xbC_{\bx_k}\}}{\mathrm{max.}} \ &  \underset{k=1,\cdots,K}{\mathrm{min.}}\   \frac{1}{2\alpha_k}\xlog\frac{1-C_{y_k}^{-2}|\xC_{y_k}|^2}{1-C_{s_k}^{-2}|\xC_{s_k}|^2}\ \notag \\
 \text{s.t.} \ &\bigg[\begin{matrix}\mathbf t_k\mathbf t_k^{H} & \xbC_{\bx_k} \\ \xbC_{\bx_k}^* & (\mathbf t_k \mathbf t_k^{H})^* \end{matrix} \bigg]\succeq \mathbf{0}, \label{C:PSD} \ \forall k.
\end{align}
 (P1.4) is a matrix optimization problem that deals with  the additional rate term in \eqref{E:Rk} due to the use of improper Gaussian signaling.
Denote the optimal value of (P1.4) as $\tau^{\star}$. If $\tau^{\star}>0$, then a strict rate improvement corresponding to rate-profile $\boldsymbol \alpha$ over the optimal proper Gaussian signaling is achieved.  
The following result will be used for solving (P1.4).
\begin{lemma}\label{T:reduceOptVar}
The positive semidefinite constraint in \eqref{C:PSD} is satisfied if and only if
\begin{align}
\xbC_{\bx_k}=Z_k\widetilde{\mathbf{t}}_k\widetilde{\mathbf{t}}_k^T, \ \forall k, \label{E:reduceOptVar}
\end{align}
where $Z_k$ is a complex scalar variable satisfying $|Z_k|\leq \|\mathbf{t}_k\|^2$, and $\widetilde{\mathbf{t}}_k=\mathbf{t}_k/\|\mathbf{t}_k\|$ denotes the normalized transmit beamforming vector for user $k$ with proper Gaussian signaling.
\end{lemma}
\begin{IEEEproof}
Please refer to Appendix~\ref{A:reduceOptVar}.
\end{IEEEproof}
It is noted that $\xbC_{\bx_k}$  in \eqref{E:reduceOptVar} is a rank-1 symmetric matrix, and thus Lemma~\ref{T:reduceOptVar} establishes the optimality of rank-1 pseudo-covariance matrices in the $K$-user MISO-IC, if the optimal rank-1 transmit covariance matrices are applied.  Furthermore, it follows from \eqref{E:reduceOptVar} that each of such rank-1 pseudo-covariance matrices is parameterized by one unknown  complex scalar $Z_k$. As a result, the problem dimension for pseudo-covariance optimization can be significantly reduced from $KM^2$ in the original matrix problem (P1.4) to $K$ by applying Lemma~\ref{T:reduceOptVar}, as will be shown next.

By substituting \eqref{E:reduceOptVar} into \eqref{E:MISO_PCyk} and \eqref{E:MISO_PCsk}, we have
\begin{align}
\xC_{y_k}=\sum_{j=1}^K (\mathbf h_{kj} \widetilde{\mathbf t}_j)^2 Z_j, \ \xC_{s_k}=\sum_{j\neq k} (\mathbf h_{kj} \widetilde{\mathbf t}_j)^2 Z_j,\ \forall k. \label{E:PCskReduced}
\end{align}
Define the following $K$-dimensional complex-valued vectors:
\begin{align}
&\mathbf z=\left[\begin{matrix}Z_1 & \cdots &  Z_K\end{matrix}\right]^T,\notag \\
& \mathbf m_k= C_{y_k}^{-1}\left[\begin{matrix}(\mathbf h_{k1} \widetilde{\mathbf t}_1)^2 & \cdots & (\mathbf h_{kK} \widetilde{\mathbf t}_K)^2\end{matrix}\right]^H, \notag \\
&\mathbf w_k = C_{s_k}^{-1} \left[\begin{matrix}\cdots &  \hspace{-1ex} (\mathbf h_{k(k-1)} \widetilde{\mathbf t}_{k-1})^2 &   0 &   (\mathbf h_{k{(k+1)}} \widetilde{\mathbf t}_{k+1})^2 & \hspace{-1ex} \cdots\end{matrix}\right]^H.\notag
  \end{align}
   Then we have
\begin{align}
&C_{y_k}^{-2}\big|\xC_{y_k} \big|^2=|\mathbf m_k^H \mathbf z|^2, \\
&C_{s_k}^{-2}\big|\xC_{s_k}\big|^2=|\mathbf w_k^H \mathbf z|^2.\label{E:PCSkSquare}
\end{align}
 Therefore, (P1.4) can be reformulated as
\begin{align}
 \text{(P1.5):} \    \underset{\mathbf z\in \mathbb{C}^K}{\mathrm{max.}} \ &  \underset{k=1,\cdots,K}{\mathrm{min.}}  \ \frac{1}{2\alpha_k}\xlog\frac{1-|\mathbf m_k^H \mathbf z|^2}{1-|\mathbf w_k^H \mathbf z|^2}\ \notag \\
 \text{s.t.} \ &|\mathbf{e}_k^T\mathbf{z}  |^2 \leq \big  \|\mathbf{t}_k \big \|^4, \ \forall k,\label{C:zjAbs}
\end{align}
where $\mathbf e_k$ is the $k$th column of a $K\times K$ identity matrix. Note that the constraint in \eqref{C:zjAbs} corresponds to the condition $|Z_k|\leq \|\mathbf t_k\|^2$ given in Lemma~\ref{T:reduceOptVar}. Before solving (P1.5), we first give an intuitive discussion on when it is possible for (P1.5) to have a strictly positive objective value, or in other words, to achieve a strict rate improvement over the optimal proper Gaussian signaling by further optimizing the pseudo-covariance matrices.

Denote the second term of the rate expression in \eqref{E:Rk} as $\Delta R_k$, which is the additional rate gain due to the use of improper Gaussian signaling. 
With the rank-1 transmit covariance and pseudo-covariance matrices given above, a simple upper bound on $\Delta R_k$ in (P1.5) can be obtained as
\begin{align}
\Delta R_k &=\frac{1}{2}\xlog\frac{1-|\mathbf m_k^H \mathbf z|^2}{1-|\mathbf w_k^H \mathbf z|^2}\\
&\leq \frac{1}{2}\xlog\frac{1}{1-|\mathbf w_k^H \mathbf z|^2}\\
&\leq -\frac{1}{2} \xlog \big(1-\|\mathbf w_k\|^2\|\mathbf z\|^2\big),\label{E:upperBound}
\end{align}
where in order for the upper bound in \eqref{E:upperBound} to be meaningful, $\|\mathbf w_k\|^2$ is assumed to be small enough so that $\|\mathbf w_k\|^2\|\mathbf z\|^2< 1$. Note that $\|\mathbf w_k\|^2=C_{s_k}^{-2}\sum_{j\neq k}|\mathbf h_{kj} \widetilde{\mathbf t}_j|^4$, which reflects the interference level experienced at receiver $k$ given the beamforming vectors $\{\mathbf t_k\}$. As $\|\mathbf w_k\|^2\rightarrow 0$, the upper bound in \eqref{E:upperBound} approaches to $0$, which leads to the following remark.
\begin{remark}\label{R:ZF}
Improper Gaussian signaling is more advantageous than proper Gaussian signaling in MISO-IC  only when there is non-negligible interference among the users. For example, with zero-forcing (ZF) transmit beamforming vectors $\{\mathbf t_{j}^{\text{ZF}}\}$, i.e., $\mathbf h_{kj} \mathbf t_{j}^{\text{ZF}}=0$, $\forall j\neq k$, we have $\|\mathbf w_k\|^2=0$ and thus $\Delta R_k= 0$, $\forall k$. As a result, no rate improvement can be achieved by further optimizing the pseudo-covariance matrices. This is as expected since ZF transmit beamforming essentially results in $K$ independent point-to-point MISO channels, where proper Gaussian signaling is known to be optimal \cite{36}.
\end{remark}
For the general $K$-user MISO-IC, ZF transmit beamforming is  feasible only when the number of transmit antennas at each transmitter is no smaller  than the total number of users, i.e., $M\geq K$. For the general case where the users are interfered by  each other with non-ZF transmit beamforming, i.e., $\mathbf w_k\neq \mathbf 0$, there is then a potential opportunity  to improve the achievable rates over the optimal proper Gaussian signaling by solving the pseudo-covariance optimization problem in (P1.5), as we pursue next.

 (P1.5) is a problem of maximizing the minimum of $K$ weighted log-fractions of quadratic functions over $\mathbf z$, for which the well-known SDR technique can be applied to find an efficient approximate solution \cite{349}.  Let $\mathbf M_k=\mathbf m_k \mathbf m_k^H$ and $\mathbf W_k=\mathbf w_k \mathbf w_k^H, \ \forall k$, and with the identity $\mathbf x^H \mathbf A \mathbf x=\xTr(\mathbf A \mathbf x \mathbf x^H)$, the SDR problem of (P1.5) is formulated as
\begin{align}
\text{(P1.5-SDR):} \    \underset{\mathbf Z\succeq \mathbf 0}{\mathrm{max.}}\ & \underset{k=1,\cdots,K}{\mathrm{min.}}\   \frac{1}{2\alpha_k}\xlog\frac{1-\mathbf \xTr(\mathbf M_k \mathbf Z)}{1-\xTr(\mathbf W_k \mathbf Z)}\ \notag \\
 \text{s.t.} \ &\xTr(\mathbf{E}_k\mathbf{Z}) \leq \big  \|\mathbf{t}_k \big \|^4, \ \forall k, \label{C:ZUpper}
\end{align}
where $\mathbf E_k=\mathbf e_k \mathbf e_k^T$. It is easy to see that (P1.5) is equivalent to (P1.5-SDR) with the additional constraint $\mathrm{rank}(\mathbf Z)=1$, under which $\mathbf Z$ can be written as $\mathbf Z=\mathbf z \mathbf z^H$. Therefore,  the optimal value of (P1.5-SDR) provides an upper bound on that of  (P1.5). 
By introducing a slack variable $\tau$, (P1.5-SDR) can be recast as
\begin{align}
   \text{(P1.5-SDR$^\prime$):} \ &  \underset{\mathbf Z\succeq \mathbf 0, \tau}{\mathrm{max.}}\   \tau \notag \\
      & \  \text{s.t.} \quad \frac{1}{2\alpha_k}\xlog\frac{1-\mathbf \xTr(\mathbf M_k \mathbf Z)}{1-\xTr(\mathbf W_k \mathbf Z)}\geq \tau, \ \forall k, \notag \\
  & \hspace{6ex} \xTr(\mathbf{E}_k\mathbf{Z}) \leq \big  \|\mathbf{t}_k \big \|^4, \ \forall k. \label{C:ZUpper}
\end{align}

 \begin{theorem}\label{T:strictPos}
For any matrix $\mathbf Z\succeq \mathbf 0$ satisfying \eqref{C:ZUpper}, the following inequalities hold:
\begin{align}
&1-\xTr(\mathbf W_k \mathbf Z)\geq C_{s_k}^{-2}\sigma^4>0, \ \forall k, \label{E:strictPos1}\\
 &1-\xTr(\mathbf M_k \mathbf Z)\geq C_{y_k}^{-2}\sigma^4>0,\ \forall k. \label{E:strictPos2}
\end{align}
\end{theorem}
\begin{IEEEproof}
Please refer to Appendix~\ref{A:strictPos}.
\end{IEEEproof}
With the inequalities in \eqref{E:strictPos1} and \eqref{E:strictPos2}, it then follows  that (P1.5-SDR$^\prime$) is a quasi-convex problem \cite{202}. For any given $\tau\geq 0$, we consider the following problem.
\begin{align}
  \text{(P1.6):}\   \underset{\mathbf Z\succeq \mathbf 0}{\mathrm{min.}}& \ \xTr(\mathbf E_1 \mathbf Z)\notag \\
\text{s.t.} \ &  1-\mathbf \xTr(\mathbf M_k \mathbf Z)\geq 2^{2\alpha_k \tau}(1-\xTr(\mathbf W_k \mathbf Z)),\ \forall k, \notag \\
  &\xTr(\mathbf{E}_k\mathbf{Z}) \leq \big  \|\mathbf{t}_k \big \|^4, \ k=2, \cdots, K. \notag
\end{align}
(P1.6) is a semidefinite programming (SDP), which minimizes  the left hand side (LHS) of \eqref{C:ZUpper} corresponding to $k=1$, subject to a target objective value $\tau$ for (P1.5-SDR$^\prime$). Denote the optimal value of (P1.6) as $f(\tau)$. If $f(\tau)\leq \|\mathbf t_1\|^4$, then the optimal value $\tau_{\text{sdr}}$  of (P1.5-SDR$^\prime$) satisfies $\tau_{\text{sdr}}\geq \tau$; otherwise, $\tau_{\text{sdr}}<\tau$. Therefore, (P1.5-SDR$^\prime$) can be optimally solved by solving the SDP problem (P1.6) with different values of $\tau$, and applying a bisection search over $\tau$.

Denote the solution to (P1.5-SDR$^\prime$) by $\mathbf Z^{\star}$. If $\mathrm{rank}(\mathbf Z^{\star})=1$, i.e., $\mathbf Z^{\star}=\mathbf z \mathbf z^H$, then $\mathbf z$ is the optimal solution to (P1.5). In this case, our proposed SDR is tight; otherwise, if $\mathrm{rank}(\mathbf Z^{\star})>1$, we then apply the following Gaussian randomization procedure customized to our problem to find an approximate solution to (P1.5) \cite{349}.
\begin{algorithm}[H]
\caption{Gaussian Randomization Method for (P1.5)}
\label{A:randomization}
\textbf{Input:} The solution $\mathbf Z^{\star}$ to (P1.5-SDR$^\prime$), and the number of randomization trials $L$.
\begin{algorithmic}[1]
\FOR{$l=1,\cdots, L$}
 \STATE Generate $\boldsymbol \xi_l\sim \mathcal{CN}(\mathbf 0, \mathbf Z^{\star})$, and construct a feasible point $\mathbf z_l$ to (P1.5) as follows:
 \begin{align}
 [\mathbf z_l]_k=\kappa_k [\boldsymbol \xi_l]_k,  \text{with } \kappa_k=\min \Big \{1, \frac{\|\mathbf t_k\|^2}{|[\boldsymbol \xi_l]_k|}\Big \}, \forall k. \notag
 \end{align}
\ENDFOR
\STATE determine $l^{\star}=\arg \underset{l=1,\cdots,L}{\mathrm{max.}}\ \underset{k=1,\cdots,K}{\mathrm{min.}} \frac{1}{2\alpha_k}\xlog\frac{1-\mathbf z_l^H \mathbf M_k \mathbf z_l}{1-\mathbf z_l^H \mathbf W_k \mathbf z_l}$
\end{algorithmic}
\textbf{Output:} $\mathbf {\hat{z}}=\mathbf z_{l^\star}$ as an approximate solution for (P1.5).
\end{algorithm}
To summarize, our proposed separate covariance and pseudo-covariance optimization algorithm for (P1) is given in Algorithm~\ref{A:P1-MISO} below.
\begin{algorithm}[H]
\caption{Separate Covariance and Pseudo-Covariance Optimization for (P1)}
\label{A:P1-MISO}
\begin{algorithmic}[1]
\STATE Solve (P1.1), denote the solution as $\{r^{\star}, \bC_{\bx_k}^{\star}=\mathbf t_k \mathbf t_k^H\}$.
\STATE  Solve (P1.5-SDR$^\prime$) using bisection search method over (P1.6), and denote the solution as $\mathbf Z^{\star}$.
\STATE If $\mathrm{rank}(\mathbf Z^{\star})=1$, then its principal component $\mathbf z$ with $\mathbf Z^{\star}=\mathbf z \mathbf z^H$ is the optimal solution to (P1.5); otherwise, find an approximate solution $\mathbf {\hat z}$ using Algorithm~\ref{A:randomization}.
\STATE Obtain the pseudo-covariance matrix solution $\{\xbC_{\bx_k}^{\star}\}$ using \eqref{E:reduceOptVar}, and the maximum sum-rate $R^{\star}$ using \eqref{E:Rk}.
\end{algorithmic}
\end{algorithm}
 \begin{remark}\label{R:optK2}
 In the special case of $K=2$, (P1.6) is a complex-valued SDP problem with three affine constraints.
 It is known  that if such a problem is feasible, there is always a rank-1 optimal solution \cite{352}. Therefore, for the two-user MISO-IC with rank-1 transmit covariance matrices, the SDR-based solution will give the optimal pseudo-covariance matrices of  rank-1. Note that in \cite{350}, a SOCP-based algorithm has been proposed, which is able to find the optimal pseudo-covariances for the two-user SISO-IC. However, the SOCP-based algorithm is difficult to be extended to the general case of $K\geq 2$ and MISO-IC with $M>1$.
 \end{remark}

 \section{Improper Gaussian Signaling for MISO-BC}\label{S:MISO-BC}
 The improper Gaussian signaling scheme discussed in the preceding sections can also be  applied to the Gaussian MISO-BC under the assumption of widely linear precoding being employed at the base station (BS) transmitter.\footnote{Widely linear precoding is more general than the conventional (strictly) linear precoding, and its resulting transmitted signals are improper in general \cite{350}.} Consider a single-cell MISO-BC where the BS with $M>1$ transmit antennas sends independent information to $K$ single-antenna receivers. The signal received by the $k$th user can be written as
 \begin{align}
 y_k&=\mathbf h_k \mathbf x + v_k \notag \\
 &=\mathbf h_k \mathbf x_k +\sum_{j\neq k} \mathbf h_k \mathbf x_j +v_k,\ k=1,\cdots,K,\label{E:MISO-BC}
 \end{align}
 where $\mathbf h_k \in \mathbb{C}^{1\times M}$ denotes the channel vector from the transmitter  to user $k$; $v_k \sim \mathcal{CN}(0, \sigma^2)$ represents the CSCG noise; and $\mathbf x=\sum_{k=1}^K \mathbf x_k$ is the  transmitted signal vector from the transmitter  with $\mathbf x_k$ denoting the transmitted signal intended for user $k$.

 It is known that the capacity of the MISO-BC  can be achieved by employing the ``dirty paper coding (DPC)'' technique  at the transmitter  \cite{212}. DPC is a nonlinear precoding  technique by which  the information for different users is encoded in a sequential manner, so that the interference caused by earlier encoded users can be completely removed at the transmitter with  non-causal information of the earlier encoded users' data streams. Since the nonlinear capacity-achieving DPC scheme is  difficult to implement in practical  systems, a great deal of research has focused on  simpler linear precoding  schemes at the transmitter. In this case, all the inter-user interferences are treated as additive Gaussian noise, thus resembling a MISO-IC (but with a transmitter's sum-power constraint that replaces the transmitters' individual power constraints in the general MISO-IC defined in Section~\ref{S:systemModel}). Denote by $\mathbf t_k$ the transmit beamforming vector for user $k$ in a $K$-user MISO-BC with conventional linear precoding, we then have
 \begin{align}
 \mathbf x_k=\mathbf t_k d_k,  \ \forall k, \label{E:xk}
 \end{align}
 where $d_k$ is a CSCG random variable representing the information signal of user $k$, i.e., $d_k\sim \mathcal{CN}(0, 1)$. The signal vector $\mathbf x_k$  in \eqref{E:xk} is a proper Gaussian RV with covariance  and pseudo-covariance matrices respectively given by
 \begin{align}
\bC_{\bx_k}=\mathbf t_k \mathbf t_k^H, \ \xbC_{\bx_k}=\mathbf 0, \ \forall k.
 \end{align}
 However, it is not immediately clear from \eqref{E:MISO-BC} whether the restriction of proper Gaussian signaling, or zero pseudo-covariance matrices of the transmitted signals, will  incur any  rate loss in the MISO-BC. Therefore, we consider the more general improper Gaussian signaling similar to the MISO-IC, where the pseudo-covariance matrices $\{\xbC_{\bx_k}\}$  are further optimized, i.e., the more general widely linear precoding \cite{350} is considered. Similar to \eqref{E:Rk}, the achievable rate in MISO-BC with improper Gaussian signaling  can be expressed as
 \begin{align}
 R_{k,\text{BC}}&=\underbrace{\xlog\left(1+  \frac{|\bh_{k}\mathbf t_k|^2}{\sigma^2+\sum_{j\neq k}|\mathbf h_k \mathbf t_j|^2} \right)}
 _{\triangleq R_{k,\text{BC}}^{\text{proper}}(\{\mathbf t_j\})}\notag \\
&\qquad \qquad \qquad
\qquad+\frac{1}{2}\xlog\frac{1-C_{y_k}^{-2}|\xC_{y_k}|^2}{1-C_{s_k}^{-2}|\xC_{s_k}|^2},\label{E:Rk-BC}
 \end{align}
 where $C_{y_k}$, $\xC_{y_k}$, $C_{s_k}$, and $\xC_{s_k}$ are defined similarly as in \eqref{E:MISO_Cyk}-\eqref{E:MISO_PCsk}. Note that the rate expression \eqref{E:Rk-BC} bears the same structure as \eqref{E:Rk} for the MISO-IC. For a given rate-profile {\boldmath$\alpha$}$=(\alpha_1\cdots \alpha_K)$, the problem of finding a Pareto-optimal rate-tuple for the MISO-BC  can thus be formulated as
\begin{align}
\text{(P1-BC):}    &  \underset{\{\mathbf t_k\},\{\xbC_{\bx_k}\}, R}{{\mathrm{max.}}}   R \notag \\
 &  \qquad \ \text{s.t.} \quad     R_{k, \text{BC}}\geq \alpha_k R, \ \forall k, \notag \\
& \qquad \qquad \ \ \bigg[\begin{matrix}\mathbf t_k\mathbf t_k^H & \xbC_{\bx_k} \\ \xbC_{\bx_k}^* & (\mathbf t_k\mathbf t_k^H)^* \end{matrix} \bigg]\succeq \mathbf{0}, \ \forall k, \label{C:validPairBC}\\
& \qquad \qquad \ \ \sum_{k=1}^K \|\mathbf t_k\|^2\leq P,\label{C:sumPower}
\end{align}
where \eqref{C:validPairBC} is the necessary and sufficient condition for $\mathbf t_k \mathbf t_k^H$ and $\xbC_{\bx_k}$ to be a valid pair of covariance and pseudo-covariance matrices, and \eqref{C:sumPower} denotes the sum-power constraint at the transmitter. Note that from the optimization perspective, (P1-BC) is identical to (P1), and hence the separate covariance and pseudo-covariance optimization algorithm presented in the previous section for the MISO-IC can be directly applied to solve (P1-BC).

\section{Numerical Results}\label{S:simulation}
In this section, we evaluate the performance of the proposed algorithm by numerical examples. All transmitters are assumed to have the same power constraint $P$, i.e., $P_k=P, \ \forall k$. The average SNR is defined as SNR$\triangleq \frac{P \mathbb{E}(|h_{kj}^m|^2)}{\sigma^2}$, where $\mathbb{E}(|h_{kj}^m|^2)$ is the average power gain from the $m$th antenna of transmitter $j$ to the antenna of receiver $k$ and is normalized to unity for all $k,j, m$. For the Gaussian randomization in Algorithm~\ref{A:randomization}, $L=1000$ is used.  
\begin{table}[htb]
\small
\caption{Mean and standard deviation (std) of the approximation ratio upper bound $\tau_{\text{sdr}}/\hat \tau$.}
\centering
\begin{tabular}{|l|l | c |c |c |c |c| c |c |c |}
\hline
 & $K$ & 2 & 3 & 4 & 5 & 6  \\
\hline
\multirow{2}{*}{$M=1$} & mean & 1.0 & 1.032 & 1.138 & 1.267 & 1.391\\
\cline{2-7}
& std & 0 &  0.092 &  0.245 & 0.350 & 0.441 \\
\hline \hline
\multirow{2}{*}{$M=2$} & mean & 1.0 & 1.012 & 1.162 & 1.401 & 1.640 \\
\cline{2-7}
& std & 0 &  0.068 &  0.388 & 0.621 & 0.691 \\
\hline
\end{tabular}
\label{Table:approxRatio}
\end{table}

\subsection{Approximation Ratio for SDR}
First, we evaluate the quality of the SDR-based approximate solution for the pseudo-covariance optimization proposed in Section~\ref{S:pseudoOpt}. Denote $\tau^{\star}$ and $\tau_{\text{sdr}}$ as the optimal objective values of (P1.5) and its relaxation (P1.5-SDR$^\prime$), respectively.  Also denote  $\hat \tau$ as the objective value of (P1.5) with the approximate solution obtained by Algorithm~\ref{A:randomization}. Then we have
\begin{align}
 \hat\tau\leq \tau^{\star} \leq\tau_{\text{sdr}},  \text{ or }  1\leq \tau^{\star}/{\hat \tau} \leq \tau_{\text{sdr}}/{\hat \tau},\notag
\end{align}
where $\tau^{\star}/{\hat \tau}$ is the approximation ratio. Since in general  the optimal value $\tau^{\star}$ is difficult to be found, the upper bound $\tau_{\text{sdr}}/{\hat \tau}$ of the approximation ratio can be used to evaluate the quality of the SDR-based solution. If $\tau_{\text{sdr}}/{\hat \tau}=1$, then the obtained SDR-based solution is in fact optimal. With the rate-profile in (P1) given as $\boldsymbol \alpha=1/K \mathbf 1$, where $\mathbf 1$ is an all-one vector, Table~\ref{Table:approxRatio} summarizes the mean and standard deviation of $\tau_{\text{sdr}}/\hat \tau$ at SNR = $10$ dB with different pairs of values for $M$ and $K$ over $1000$ random channel realizations,  each with the channel coefficients drawn from the i.i.d. CSCG random variables with zero-mean and unit-variance. 
 It is observed that for all the setups considered, the mean values of the approximation ratio upper bounds are between $1$ and $1.64$. In particular, for $K=2$, $\tau_{\text{sdr}}/\hat \tau =1$ is guaranteed, which verifies the optimality of the SDR-based pseudo-covariance optimization for the two-user MISO-IC, as discussed in Remark~\ref{R:optK2}.

 \begin{table*}[htb]
\small
\caption{Channel realizations for Figs.~\ref{F:region10dBChannel1} and \ref{F:region10dBChannel2}.}
\centering
\begin{tabular}{|c|l | l |}
\hline
 & channel realization $\mathbf H^{(1)}$& channel realization $\mathbf H^{(2)}$  \\
\hline
$\mathbf h_{11}$& $[\begin{matrix} 0.3676e^{-1.7037i}& 0.4993e^{1.6076i}
 \end{matrix}]$ & $[\begin{matrix} 0.3676e^{-1.7037i}& 0.4993e^{1.6076i}
 \end{matrix}]$\\
 \hline
 $\mathbf h_{21}$& $[\begin{matrix}0.2526e^{-1.8997i}& 0.3270e^{\mathbf{1.5884}i}
  \end{matrix}]$ & $[\begin{matrix}0.2526e^{-1.8997i}& 0.3270e^{\mathbf{-0.3810}i}
 \end{matrix}]$\\
 \hline
 $\mathbf h_{22}$& $[\begin{matrix}0.4694e^{-0.1915i}& 0.5682e^{0.5302i}
 \end{matrix}]$ & $[\begin{matrix}0.4694e^{-0.1915i}& 0.5682e^{0.5302i}
 \end{matrix}]$\\
 \hline
 $\mathbf h_{12}$& $[\begin{matrix}0.2885e^{-0.2454i}& 0.3656e^{\mathbf{0.4710}i}
 \end{matrix}]$ & $[\begin{matrix}0.2885e^{-0.2454i}& 0.3656e^{\mathbf{1.8673}i}
 \end{matrix}]$\\
\hline
$\cos\theta_1$ & $0.9961$  & $0.6601$\\
\hline
$\cos\theta_2$ & $0.9997$  & $0.7793$\\
\hline
\end{tabular}
\label{Table:channels}
\end{table*}

\begin{figure}
\centering
\includegraphics[scale=0.44]{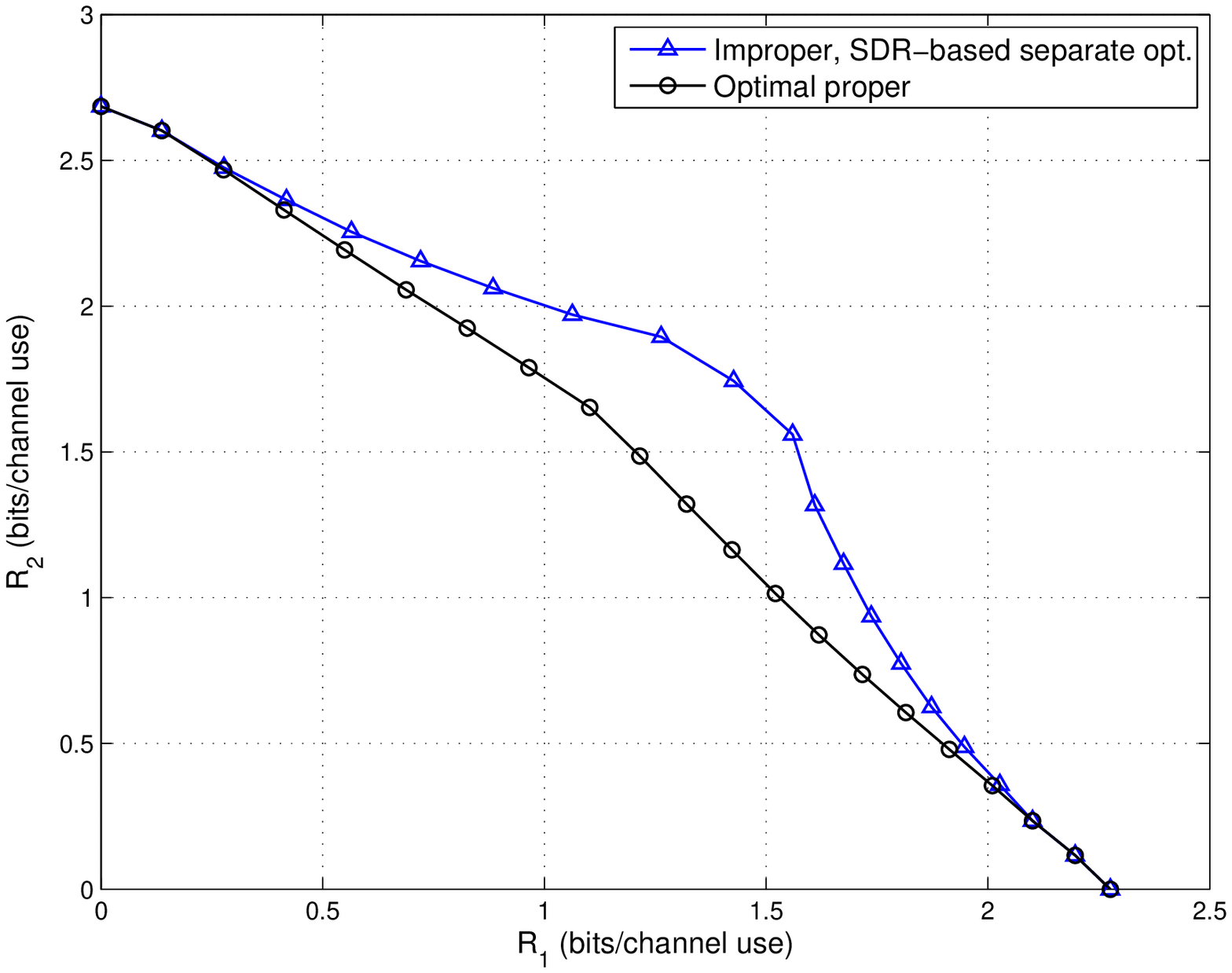}
\caption{Achievable rate region for the two-user MISO-IC with $M=2$, SNR = 10 dB, and channel realization $\mathbf H^{(1)}$.}\label{F:region10dBChannel1}
\end{figure}

\begin{figure}
\centering
\includegraphics[scale=0.44]{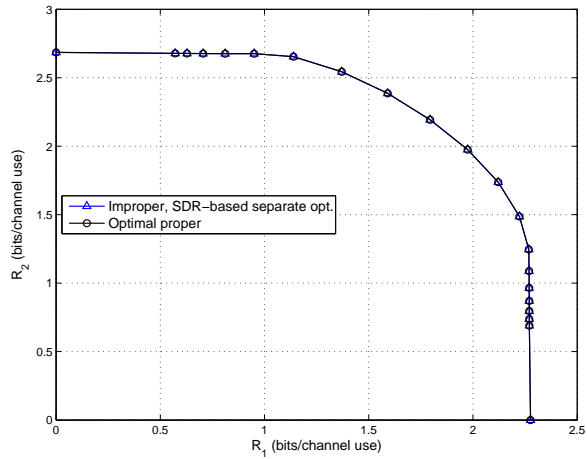}
\caption{Achievable rate region for the two-user MISO-IC with  $M=2$, SNR = 10 dB, and channel realization $\mathbf H^{(2)}$.}\label{F:region10dBChannel2}
\end{figure}

\subsection{Rate Region Comparison}\label{S:region}
In Fig.~\ref{F:region10dBChannel1}, the achievable rate region for an example two-user  MISO-IC with $M=2$ is plotted for SNR=$10$ dB, with the channel realization (denoted as $\mathbf H^{(1)}$) given in the left column of Table~\ref{Table:channels}. The proposed improper Gaussian signaling  with SDR-based separate covariance and pseudo-covariance optimization is compared against the optimal proper Gaussian signaling obtained by solving (P1.1). It is observed that for this channel setup,  the achievable rate region has been significantly enlarged by the proposed pseudo-covariance optimization for improper Gaussian signaling.

Next, we consider a different channel realization $\mathbf H^{(2)}$, which differs from $\mathbf H^{(1)}$ only in the phases of the second elements in $\mathbf h_{21}$ and $\mathbf h_{12}$, as shown in the right column of Table~\ref{Table:channels}. It is observed from Fig.~\ref{F:region10dBChannel2} that for this new channel setup, there is no notable rate gain by the proposed improper Gaussian signaling over the optimal proper signaling, which is in sharp contrast to the result in Fig.~\ref{F:region10dBChannel1}. This can be explained by comparing the residue interference levels after applying the optimal \emph{proper} Gaussian signaling in the two cases. It is worth noting that for the two-user MISO-IC, the optimal proper Gaussian signaling at each transmitter is known to be a linear combination of the ZF beamforming  and the maximum  ratio transmission (MRT) \cite{249}. Applying this result to the two-user MISO-IC in our context, it then follows that one user will generate less amount of interference to the other user if its direct channel and interfering channel vectors are more close to be orthogonal. 
Let $\theta_1$ denote the angle between the direct and interfering channel vectors for transmitter $1$, i.e., $\cos\theta_1\triangleq |\mathbf h_{21} \mathbf h_{11}^H|/(\|\mathbf h_{21}\|\|\mathbf h_{11}\|)$, and define $\theta_2$ for transmitter $2$ similarly. As shown in Table~\ref{Table:channels}, since $\theta_1$ and $\theta_2$ are smaller in the case of $\mathbf H^{(1)}$ than that in the case of $\mathbf H^{(2)}$,  more substantial interference is resulted even after applying  the optimal proper transmit beamforming. As a result,  further optimization over the pseudo-covariance matrices provides more significant rate gains in the case of $\mathbf H^{(1)}$ than that of $\mathbf H^{(2)}$, which is consistent with our discussion  in Remark~\ref{R:ZF}.

\begin{figure}
\centering
\includegraphics[scale=0.44]{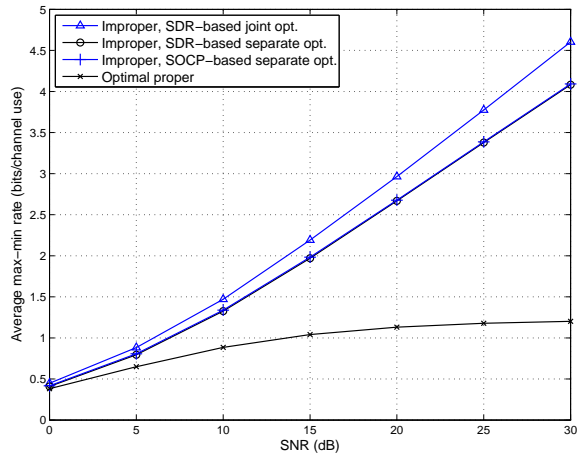}
\caption{Average max-min rate for the two-user SISO-IC.}\label{F:maxMin_K2_M1}
\end{figure}

\subsection{Max-Min Rate Comparison}
The rate-profile technique used in characterizing the Pareto boundary  of the achievable rate region can be directly applied for  maximizing the minimum (max-min) rate  of all users without time sharing. Specifically, the max-min problem for the $K$-user MISO-IC is equivalent to solving (P1) by using the rate-profile $\boldsymbol \alpha=1/K \mathbf 1$. In this subsection, the average max-min rate achievable over $1000$ random channel realizations by the proposed improper Gaussian signaling is compared with that by the optimal proper Gaussian signaling obtained by solving (P1.1).

\begin{figure}
\centering
\includegraphics[scale=0.44]{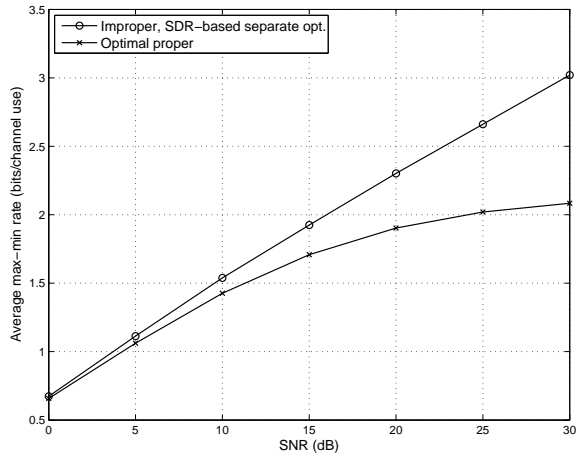}
\caption{Average max-min rate for the three-user MISO-IC with $M=2$.}\label{F:maxMin_K3_M2}
\end{figure}

We first consider the special case of the two-user SISO-IC, where two alternative improper Gaussian signaling designs proposed in \cite{350}, one with SDR-based joint covariance and pseudo-covariance optimization and the other with SOCP-based separate covariance and pseudo-covariance optimization, are compared with the proposed SDR-based separate covariance and pseudo-covariance optimization. It is observed in Fig.~\ref{F:maxMin_K2_M1} that improper Gaussian signaling provides significant rate gains over the optimal proper Gaussian signaling. At high SNR, the max-min rate by proper Gaussian signaling saturates, since the total number of data streams transmitted, which is  $2$, exceeds the total number of DoF of the two-user SISO-IC, which is known to be $1$. In contrast, a linear increase of the max-min rate with respect to the logarithm of SNR is observed by improper Gaussian signaling. 
 It is also observed that the two separate covariance and pseudo-covariance optimization algorithms,  SDR-based or SOCP-based, provide the same max-min rate performance in the case of $K=2$. This is as expected since both algorithms achieve the global optimality for the covariance and pseudo-covariance optimization subproblems in the two-user SISO-IC case. Moreover, it is observed in Fig.~\ref{F:maxMin_K2_M1} that the SDR-based joint covariance and pseudo-covariance optimization algorithm in \cite{350} achieves additional rate gains over the SDR/SOCP-based separate optimization. However, the extension of such a joint optimization to the general $K$-user MISO-IC with $K>2$ and/or $M>1$ remains unknown.

To show the max-min rate performance with improper Gaussian signaling when there are multiple transmit antennas, we consider a three-user MISO-IC with $M=2$. As shown in Fig.~\ref{F:maxMin_K3_M2}, a significant rate improvement is observed by the proposed improper Gaussian signaling over the optimal proper Gaussian signaling.

\begin{figure}
\centering
\includegraphics[scale=0.44]{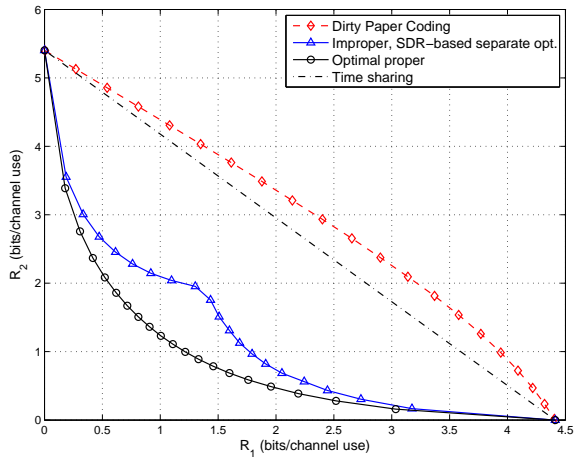}
\caption{Achievable rate region for the two-user MISO-BC with  $M=2$ and SNR = 10 dB.}\label{F:BC_Region10dB}
\end{figure}

\subsection{Rate Region of   MISO-BC}
At last,  MISO-BC with improper Gaussian signaling or widely linear precoding as discussed in Section~\ref{S:MISO-BC} is considered. In Fig.~\ref{F:BC_Region10dB}, the achievable rate regions for an example two-user MISO-BC with $M=2$ are plotted for SNR = 10 dB. The channel vectors of the two users are respectively given by $\mathbf h_1=[\begin{matrix} 1.1741e^{1.0030i}& 0.8064e^{2.8642i} \end{matrix}]$, $\mathbf h_2=[\begin{matrix}1.8116e^{2.0647i}& 0.9209e^{-2.4167i}\end{matrix}]$. The proposed improper Gaussian signaling is compared with the optimal proper Gaussian signaling with linear beamforming. As a benchmark, the capacity region achieved by the nonlinear DPC with proper Gaussian signaling  is also included. It is observed that without time sharing of the achievable users' rates by a convex-hull operation, the achievable rate region for the MISO-BC is significantly enlarged by employing improper Gaussian signaling over the optimal proper Gaussian signaling. However, such a rate  gain no longer exists if time sharing  is performed.\footnote{
 By constructing one particular channel realization, the authors in \cite{382} have recently demonstrated  that improper Gaussian signaling is able to provide better achievable rates than proper Gaussian signaling for BC with widely linear precoding even after time sharing.}

\section{Conclusion}\label{S:conclusion}
This paper has studied the transmit optimization problem for the  $K$-user MISO-IC with the interference treated as noise, when  improper or circularly asymmetric complex Gaussian signaling is applied. By exploiting the separable achievable rate structure with improper Gaussian signaling, a separate transmit covariance and pseudo-covariance optimization algorithm has been proposed. For the pseudo-covariance optimization, we have established the  optimality of rank-1 pseudo-covariance matrices, given the optimal rank-1 transmit covariance matrices obtained by existing methods. Moreover, we have shown that each rank-1 pseudo-covariance matrix is parameterized by one unknown complex scalar and thereby the complexity for searching is greatly reduced.  We then applied the SDR technique to find an efficient approximate solution for the pseudo-covariance optimization problem. The proposed improper Gaussian signaling has been extended to MISO-BC with widely linear precoding. Finally,  numerical results have been provided to demonstrate the achievable  rate gains over conventional  proper Gaussian signaling in various multiuser multi-antenna  systems that can be  modeled by the MISO-IC.
\appendices
\section{Proof of Lemma~\ref{T:reduceOptVar}}\label{A:reduceOptVar}
The following result will be used for proving Lemma~\ref{T:reduceOptVar}.
\begin{lemma} \label{L:psdBlock} \cite{202}
If $\mathbf{X}$ is Hermitian and is partitioned as $\mathbf{X}=\left[\begin{matrix}\mathbf{A} &\mathbf{B} \\ \mathbf{B}^H & \mathbf{C} \end{matrix} \right]$, then $\mathbf{X}\succeq \mathbf{0}$ if and only if the following three conditions are satisfied:
\[
(a) \ \mathbf{A}\succeq \mathbf{0}; \ (b) \ (\mathbf{I}-\mathbf{A}\mathbf{A}^\dag)\mathbf{B}=\mathbf{0}; \ (c) \  \mathbf{C}-\mathbf{B}^H\mathbf{A}^\dag\mathbf{B}\succeq \mathbf{0},
\]
where $(\cdot)^\dag$ represents the Moore-–Penrose pseudo-inverse.
\end{lemma}
We now re-express the positive semidefinite constraint in \eqref{C:PSD} using the above three conditions. First, it is clear that $(a)$ is guaranteed in \eqref{C:PSD}. Next, to use the condition in $(b)$, we express the singular value decomposition (SVD) of the rank-1 covariance matrix $\bC_{\bx_k}^{\star}$ as
\[
\bC_{\bx_k}^{\star}=\mathbf{t}_k\mathbf{t}_k^H=\left[\begin{matrix} \widetilde{\mathbf{t}}_k & \mathbf{U}_k\end{matrix}\right]
\left[\begin{matrix} \|\mathbf{t}_k\|^2 & \mathbf{0} \\ \mathbf{0} & \mathbf{0} \end{matrix} \right]\left[\begin{matrix} \widetilde{\mathbf{t}}_k^H \\ \mathbf{U}_k^H\end{matrix} \right],
\]
where $\widetilde{\mathbf{t}}_k=\mathbf t_k/\|\mathbf t_k\|$, and $\mathbf{U}_k\in \mathbb{C}^{M \times (M-1)}$ satisfies $\mathbf{U}_k^H\mathbf{U}_k=\mathbf{I}_{M-1}$, and $\widetilde{\mathbf{t}}_k ^H\mathbf{U}_k=\mathbf{0}$. Then the Moore-–Penrose pseudo-inverse of $\bC_{\bx_k}^{\star}$ can be obtained as
\[
(\bC_{\bx_k}^{\star})^\dag=\|\mathbf{t}_k\|^{-2}\widetilde{\mathbf{t}}_k\widetilde{\mathbf{t}}_k^H.
\]
Applying the condition in $(b)$ of Lemma~\ref{L:psdBlock} to \eqref{C:PSD} yields
\begin{align}
\left(\mathbf{I}-\bC_{\bx_k}^{\star}(\bC_{\bx_k}^{\star})^\dag \right)\xbC_{\bx_k}=\mathbf{0}
& \Longleftrightarrow  (\mathbf{I}-\widetilde{\mathbf{t}}_k\widetilde{\mathbf{t}}_k^H)\xbC_{\bx_k}=\mathbf{0} \notag \\
&\Longleftrightarrow  \xbC_{\bx_k}= \widetilde{\mathbf{t}}_k \mathbf{v}_k^H \label{E:CjTildePrev}\\
&\Longleftrightarrow  \xbC_{\bx_k}=\gamma_k \widetilde{\mathbf{t}}_k \mathbf{\widetilde v}_k^H,\label{E:CjTilde}
\end{align}
 where $\mathbf v_k\in \mathbb{C}^{M\times 1}$ is an arbitrary vector with its Euclidian norm  denoted by $\gamma_k$ and normalized  direction by $\mathbf{\tilde v}_k$. To show both the ``if'' and ``only if'' conditions  in \eqref{E:CjTildePrev}, we first note that $\xbC_{\bx_k}= \widetilde{\mathbf{t}}_k \mathbf{v}_k^H$ is a solution of $(\mathbf{I}-\widetilde{\mathbf{t}}_k\widetilde{\mathbf{t}}_k^H)\xbC_{\bx_k}=\mathbf{0}$. In addition, any solution
$\xbC_{\bx_k}$ of the equation $(\mathbf{I}-\widetilde{\mathbf{t}}_k\widetilde{\mathbf{t}}_k^H)\xbC_{\bx_k}=\mathbf{0}$ must satisfy $\xbC_{\bx_k}= \widetilde{\mathbf{t}}_k (\xbC_{\bx_k}^H \widetilde{\mathbf{t}}_k)^H$, which confirms  the existence of a vector $\mathbf v_k=\xbC_{\bx_k}^H \widetilde{\mathbf{t}}_k$ such that $\xbC_{\bx_k}= \widetilde{\mathbf{t}}_k \mathbf{v}_k^H$. 

 Since  $\xbC_{\bx_k}$ is a pseudo-covariance matrix, which must be symmetric,  we have
\begin{align}
\xbC_{\bx_k}=\gamma_k \widetilde{\mathbf{t}}_k \mathbf{\widetilde v}_k^H=\gamma_k\mathbf{\widetilde v}_k^* \widetilde{\mathbf{t}}_k^T=\xbC_{\bx_k}^T.\notag
\end{align}
By expressing $\xbC_{\bx_k}\xbC_{\bx_k}^H$ using two alternative forms, we have
\begin{align}
&\xbC_{\bx_k}\xbC_{\bx_k}^H=\gamma_k^2\widetilde{\mathbf{t}}_k\widetilde{\mathbf{t}}_k^H=\gamma_k^2\mathbf{\widetilde v}_k^*(\mathbf{\widetilde v}_k^*)^H \notag\\
& \Longleftrightarrow  \widetilde{\mathbf{t}}_k=e^{i\theta_k}\mathbf{\widetilde v}_k^*, \text{ or } \mathbf{\widetilde v}_k=e^{i\theta_k}\widetilde{\mathbf{t}}_k^*,
\end{align}
where $\theta_k\in [0, 2\pi)$. By substituting $\mathbf{v}_k$ into \eqref{E:CjTilde}, we have
\begin{align}
\xbC_{\bx_k}=\gamma_k e^{-i\theta_k} \widetilde{\mathbf{t}}_k \widetilde{\mathbf{t}}_k^T=Z_k \widetilde{\mathbf{t}}_k \widetilde{\mathbf{t}}_k^T,
\end{align}
where we have defined the complex variable $Z_k=\gamma_k  e^{-i\theta_k}$. Furthermore, the condition in $(c)$ of Lemma~\ref{L:psdBlock} implies that for the constraint in \eqref{C:PSD} to be satisfied, we need to have
\begin{align}
&(\bC_{\bx_k}^{\star})^*-\xbC_{\bx_k}^*(\bC_{\bx_k}^{\star})^\dag\xbC_{\bx_k}\succeq \mathbf{0}\notag\\
&\Longleftrightarrow \left(\|\mathbf{t}_k\|^2-|Z_k|^2\|\mathbf{t}_k\|^{-2}\right)\widetilde{\mathbf{t}}_k^*\widetilde{\mathbf{t}}_k^T\succeq \mathbf{0}\notag\\
&\Longleftrightarrow |Z_k|\leq \|\mathbf{t}_k\|^2.
\end{align}
This thus completes the proof of Lemma~\ref{T:reduceOptVar}.
\section{Proof of Theorem~\ref{T:strictPos}}\label{A:strictPos}
Since the inequalities in \eqref{E:strictPos1} and \eqref{E:strictPos2} of Theorem~\ref{T:strictPos} can be proved similarly, for brevity, we only show the proof of \eqref{E:strictPos1} in this appendix.
First, if $\mathbf W_k=\mathbf w_k\mathbf w_k^H=\mathbf 0$, i.e., $\mathbf w_k$ is a zero-vector, then \eqref{E:strictPos1} is satisfied since
\begin{align}
C_{s_k}^2=(\sum_{j\neq k}|\mathbf h_{kj}\mathbf t_j|^2+\sigma^2)^2\geq \sigma^4 \label{E:Cskleqsigma}
\end{align}
 is true.
Therefore, in the following, we assume without loss of generality that at least one of the elements in $\mathbf w_k$ is non-zero. Then we consider the following optimization problem.
\begin{align}
 \text{(P-A1):}\  \underset{\mathbf Z\succeq \mathbf 0}{\mathrm{max.}} \ & \xTr(\mathbf W_k \mathbf Z)=\xTr(\mathbf w_k \mathbf w_k^H \mathbf Z)\notag \\
  \text{s.t.} \ &  \xTr(\mathbf{E}_k\mathbf{Z}) \leq \big  \|\mathbf{t}_k \big \|^4, \  \forall k. \label{C:PA1}
 \end{align}
 In order to show \eqref{E:strictPos1} in Theorem~\ref{T:strictPos}, it is sufficient to prove that  the optimal objective value of (P-A1) is no greater than $1-C_{s_k}^{-2}\sigma^4$.  With \eqref{E:Cskleqsigma}, this is clearly true if the optimal solution $\mathbf Z^{\star}$ to (P-A1) is a zero matrix. Therefore, in the following, we assume $\mathbf Z^{\star}\neq \mathbf 0$. The following result, whose proof can be found in Lemma I of \cite{219} or in Lemma 1.6 of \cite{381}, will be used.

  \begin{lemma}\label{L:rank1}
  There exists a rank-1 optimal solution to (P-A1).
  \end{lemma}

 With Lemma~\ref{L:rank1}, (P-A1) can be re-expressed as
\begin{align}
\text{(P-A2):}  \  \underset{\mathbf z}{\mathrm{max.}} & \  \mathbf z^H \mathbf W_k \mathbf z \notag \\
\text{s.t.} & \ |\mathbf{e}_k^H\mathbf{z} \big  |^2 \leq \big  \|\mathbf{t}_k \big \|^4, \ \forall k.\label{C:zjAbsUpper}
\end{align}
Using \eqref{E:PCskReduced} and \eqref{E:PCSkSquare}, we have
 \begin{align}
 \mathbf z^H \mathbf W_k \mathbf z & = C_{s_k}^{-2}\Big|\sum_{j\neq k}(\mathbf h_{kj} \widetilde{\mathbf t}_j )^2Z_j \Big|^2  \notag \\
 & \leq C_{s_k}^{-2}\Big(\sum_{j\neq k} |\mathbf h_{kj} \widetilde{\mathbf t}_j |^2 |Z_j|\Big)^2 \label{E:triangleIneq} \\
 & \leq C_{s_k}^{-2}\Big(\sum_{j\neq k} |\mathbf h_{kj} \widetilde{\mathbf t}_j |^2 \|\mathbf t_j\|^2\Big)^2 \label{E:ZjUpper}\\
 & \leq C_{s_k}^{-2}\Big[\Big(\sum_{j\neq k} |\mathbf h_{kj} \mathbf t_j |^2 +\sigma^2\Big)^2-\sigma^4\Big] \label{E:NoisePos}\\
 & =1-C_{s_k}^{-2}\sigma^4. \label{E:eqTo1}
 \end{align}
 where \eqref{E:triangleIneq} is due to the triangle inequality; \eqref{E:ZjUpper} is due to the constraint in \eqref{C:zjAbsUpper}, which is equivalent to $|Z_k|\leq \|\mathbf t_k\|^2, \forall k$;  and \eqref{E:eqTo1} is true since $C_{s_k}=\sum_{j\neq k} |\mathbf h_{kj} \mathbf t_j |^2 +\sigma^2$.  The above result shows that the optimal objective value of (P-A2), and hence that of (P-A1), is no larger  than $1-C_{s_k}^{-2}\sigma^4$, as desired.

 This thus completes the proof of Theorem~\ref{T:strictPos}.
 \bibliographystyle{IEEEtran}
\bibliography{IEEEabrv,IEEEfull}
\end{document}